\documentclass[fleqn,10pt]{wlscirep}

\usepackage[caption=false]{subfig}
\usepackage{ifpdf}
\ifpdf
\usepackage{graphicx}

\usepackage{epstopdf}
\usepackage{xfrac}
\usepackage{lmodern}
\usepackage{colortbl}

\AppendGraphicsExtensions{.tif}
\else
\usepackage{graphicx}
\fi
\graphicspath{{./figures/}}

\author[1]{Paul C. Spruijtenburg}

\author[1]{Sergey V. Amitonov}
\author[1]{Filipp Mueller}
\author[1]{Wilfred G. van der Wiel}
\author[1,*]{Floris A. Zwanenburg}
\affil[1]{NanoElectronics Group, MESA+ Institute for Nanotechnology, University of Twente, P.O. Box 217, 7500 AE Enschede The Netherlands}

\affil[*]{f.a.zwanenburg@utwente.nl}

\title{Passivation and characterization of charge defects in ambipolar silicon quantum dots}

\date{\today}

\begin{abstract}
	In this Report we show the role of charge defects in the context of the formation of electrostatically defined quantum dots. We introduce a barrier array structure to probe defects at multiple locations in a single device. We measure samples both before and after an annealing process which uses an $\textrm{Al}_2\textrm{O}_3$ overlayer, grown by atomic layer deposition. After passivation of the majority of charge defects with annealing we can electrostatically define hole quantum dots up to 180~nm in length. Our ambipolar structures reveal amphoteric charge defects that remain after annealing with charging energies of $\simeq$10~meV in both the positive and negative charge state. 
\end{abstract}

\begin{document}

\flushbottom
\maketitle

In order to perform sufficient coherent operations in a proposed quantum computer \cite{Ladd2010}, the quantum states of the corresponding qubits are required to be long-lived. In the scheme proposed by Loss and DiVincenzo\cite{Loss1998}, quantum logic gates perform operations on coupled spin states of single electrons in neighboring quantum dots (QDs).  Most experiments have focused on quantum dots formed in III-V semiconductors, especially GaAs \cite{RevModPhys.75.1,Hanson2007}. However, electron spin coherence in those materials is limited by hyperfine interactions with nuclear spins and spin-orbit coupling. Group IV materials have been shown to have long spin lifetimes because of weak spin-orbit interactions and the predominance of spin-zero nuclei. This prospect has stimulated significant experimental effort to isolate single charges in carbon nanotubes \cite{Jarillo-Herrero2004, Kuemmeth:2014uh}, Si/SiGe heterostructures \cite{Simmons2009, Borselli2014}, Si nanowires \cite{Zwanenburg2009}, planar Si MOS structures \cite{Lim:2009uz}, and dopants in Si \cite{Lansbergen2008, vanderHeijden:2014fp, Fuechsle2012, Prati2012a,Pierre2010}. Silicon not only holds promise for very long coherence times \cite{Steger2012,Veldhorst2014,Veldhorst2015}, but also for bringing scalability of quantum devices one step closer, and has thus attracted much attention for quantum computing purposes \cite{Morton2011,Zwanenburg2013}.
For the quantum states to be well-defined, it is necessary to define singular quantum dots. Disorder in the confinement potential can lead to unintentional quantum dots\cite{Spruijtenburg1,Li2013}, resonant tunneling features\cite{Jovanovic1994,Hofheinz:2006eo}, and Anderson localization\cite{Prati2012}.
In silicon planar quantum dots, nanometer-size Coulomb islands are electrostatically defined in silicon in a gated MOSFET-type heterostructure. Apart from the quality of the crystalline silicon the quality and disorder in the heterostructure is directly linked to the disorder of the confinement potential.

 In this Report, we focus on disorder caused by defects at the Si/$\textrm{Si}\textrm{O}_2$ interface and investigate their effect on the formation of electrostatically defined quantum dots. We do this by creating an array of barrier electrodes, each capable of locally controlling the potential for quantum dot formation and probing the respective crystalline environment. We show that an annealing process with $\textrm{Al}_2\textrm{O}_3$, grown by atomic layer deposition (ALD), reduces the disorder such that quantum dots 180~nm in length can be formed. Furthermore, we observe that after annealing a small percentage of barriers show charge transitions with an amphoteric character.
% to form tunnel barriers for a quantum dot, or to.

% We focus our attention on the role of defects formed at the interface between Si and $\textrm{Si}\textrm{O}_2$ during oxide growth and their influence on the formation of electrostatically defined quantum dots.

The role of defects at the Si/$\textrm{Si}\textrm{O}_2$ interface in this heterostructure has been extensively studied in the context of MOSFET technology. One of the most studied defects at this interface is the paramagnetic $\textrm{P}_\textrm{b}$ center\cite{Nishi1972,Thoan2011,Mishima2000} which is linked to surface charges, decreased mobility, and the negative-bias temperature instability (NBTI) effect\cite{Schroder2003,Campbell2007}. For quantum computation, $\textrm{P}_\textrm{b}$ centers have been used in coherent manipulation of spin-dependent charge-carrier recombination of phosphorus donors.\cite{Stegner2006} Defects at the interface have also been associated with random telegraph signals at low temperatures and characterized by means of single-electron spin resonance \cite{Xiao2003, Xiao2004} and magnetic-field dependent measurements\cite{Prati2006}.

The $\textrm{P}_\textrm{b}$ center is in essence a Si atom with a dangling bond that has formed due to the incommensurability of the crystalline Si and the amorphous $\textrm{Si}\textrm{O}_2$. Two types of $\textrm{P}_\textrm{b}$ center can be further distinguished. The $\textrm{P}_\textrm{b0}$ center is backbonded to two Si atoms and one oxygen atom, while the $\textrm{P}_\textrm{b1}$ center is backbonded to three Si atoms.
For the $\textrm{P}_\textrm{b0}$ center, the density of interface states D$_{\text{it}}$ in the gap is amphoteric in nature with its maxima at $0.25$ and $0.85$ eV above the valence band \cite{Ragnarsson2000}.  
%For an isolated $\textrm{P}_\textrm{b}$ center however, the D$_{\text{it}}$ picture breaks down. Instead, it is more accurate to describe it as a single dopant with discrete levels in the bandgap. 
%In intrinsic silicon the $\textrm{P}_\textrm{b}$ center is expected to have three charge states: positive $\textrm{D}^\textrm{+}$, neutral $\textrm{D}^\textrm{0}$, and negative $\textrm{D}^\textrm{-}$ with 0, 1, or 2 electrons on the $\textrm{P}_\textrm{b}$ center, respectively\cite{Lannoo1990}.
Passivating $\textrm{P}_\textrm{b}$ centers is commonly achieved by introducing hydrogen to the dangling bonds in an annealing process. This process reduces the dangling bond spatial density from order $10^{13}~\text{cm}^{-2}$  to $10^{10}~\text{cm}^{-2}$ and improves numerous parameters related to MOSFET-like transistor operation and has also been shown to improve parameters related to quantum dot formation \cite{Cartier1993,Angus2007}.

\begin{figure}
	\centering
	\includegraphics[width=\linewidth]{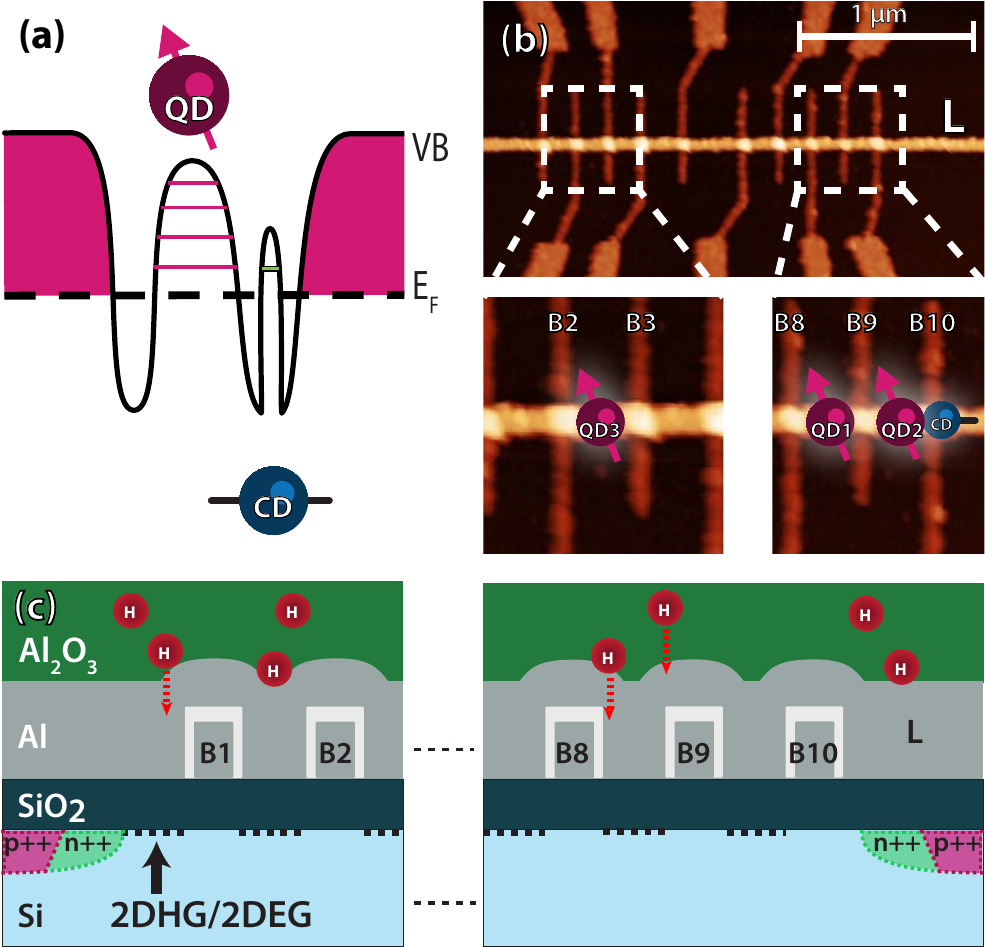}
\begin{minipage}{.5\columnwidth}
	% \Mybox{
	\subfloat{
	\raggedleft
	\label{fig:schematic_potentialprofile}
	}%}
\end{minipage}%
\begin{minipage}{.5\columnwidth}
	\centering
	\subfloat{
	\label{fig:AFM}
	}%
	\subfloat{
	\label{fig:schematic_layout}
	}
\end{minipage}%
	\caption[LoF]{Si barrier array device structure. \protect\subref{fig:schematic_potentialprofile} Electrochemical potential profile of a hole quantum dot with and without a passivated and unpassivated charge defect in the left and right tunnel barrier respectively. In case of an unpassivated charge defect (``CD''), an extra level in the bandgap (green) is available for transport. 
\protect\subref{fig:AFM} Atomic force microscopy image of the device, showing the lead gate L horizontally across the image. The barriers B1 through B10 allow local probing of the 2DHG/2DEG at many locations.  \protect\subref{fig:schematic_layout} A schematic cross-sectional image of the device. The hydrogen (H) present in the $\textrm{Al}_2\textrm{O}_3$ can passivate charge defects during annealing. The applied voltage on the aluminum gates creates a 2DHG/2DEG, indicated by the dashed lines.
}

\label{fig:one}
\end{figure}

Addressing only a single quantum dot defined by electrostatic gating can be hampered by a charge defect. If a charge defect is located beneath an electrode intended to form a tunnel barrier, this electrode acts a gate on this defect. 
Instead of forming a single tunnel barrier to the QD, the levels of the defect remain available for transport resulting in additional single-charge tunneling events.
This situation is sketched for an unpassivated charge defect in Figure~\ref{fig:schematic_potentialprofile} for the right tunnel barrier.

While the valence band of the right tunnel barrier is below the Fermi level, the level of the charge defect is above the Fermi level and is thus available for hole transport. If the defect is in close enough proximity to the QD, an ill-defined double quantum dot can form. If the defect is passivated by hydrogen, the potential is smooth and behaves as a normal tunnel barrier for the QD. The same arguments hold for an impurity level just below the conduction band when measuring an electron QD.
% without any disruptions

Figure~\ref{fig:AFM} shows an atomic force microscope image and Figure~\ref{fig:schematic_layout} a schematic cross section of our device structure, made with a combination of optical and electron-beam lithography (EBL), based on the recipe as described by Angus \emph{et al.}\cite{Angus2007}. We further distinguish two device types --- A and B --- specified below.
Near-intrinsic (100) silicon ($\rho \geq 10~\text{k}\Omega\cdot\text{cm}$) is used as the substrate. Source and drain regions are implanted with boron/phosphorus dopant atoms, which are activated by rapid thermal annealing, and serve as hole/electron reservoirs. Ohmic contacts to these regions are made by sputtering  Al-Si alloy (99:1) contact pads. A 10~nm thick high-quality $\textrm{Si}\textrm{O}_2$ oxide window is thermally grown at $900^\circ\text{C}$ and serves as an insulating barrier between the substrate and the aluminum gates. Contact pads for gates are defined using optical lithography followed by development, evaporation of Ti/Pd, and subsequent lift-off. At this point, for device type B, a $5$ nm $\textrm{Al}_2\textrm{O}_3$ ``underlayer'' using ALD at a process temperature of $250^\circ\text{C}$ is grown on top of the $\textrm{Si}\textrm{O}_2$.
Subsequently, EBL followed by evaporation and liftoff is used to define the sub-micron aluminum gates. These gates will electrostatically control charge accumulation. In the first step, the barrier gates are created and thermally oxidized. In the second step, the lead gate is created.
Atomic force microscopy images show barrier gates with a typical width of 35~nm, and a minimum separation of 180~nm, while the lead gate has a width of approximately 80~nm, see Figure~\ref{fig:AFM}.
Following the creation of the aluminum gate structures, we perform a first set of measurements. Then we grow a final ``overlayer'' of 5~nm $\textrm{Al}_2\textrm{O}_3$ at $100^\circ\text{C}$, and the sample is heated to $300^\circ\text{C}$ for 45 minutes in pure Ar ambient of 10~mbar. As a final step, the sample is cleaned in UV ozone for 2 minutes to remove any moisture present.

% The ambipolar device structure\cite{Mueller2015} contains spatially separated highly p-type/n-type doped source and drain regions and is shown in the intrinsic silicon, on top of which is the SiO$_2$ barrier. The Al gates are evaporated on top and electrically isolated from each other by aluminum oxide. The ALD deposited $\textrm{Al}_2\textrm{O}_3$ containing hydrogen is shown in green as the top layer.
%we use a twofold approach
%we can locally probe
To study both defects and their influence on QD formation, we use a twofold approach: by using an array of electrodes (seen in an AFM image in Figure~\ref{fig:AFM}), each acting as a single barrier, we can locally probe the 2D carrier gas and observe single-charge tunneling events. At the same time, using an ambipolar device structure\cite{Mueller2015,Betz2014} it is possible to infer if accessible levels are located near the conduction or valence band edge by either inducing a 2D hole gas (2DHG) or 2D electron gas (2DEG). Thus, our ambipolar structures allow investigation of both the hole and electron nature of these defects. 

Recent work has shown that $\textrm{Al}_2\textrm{O}_3$ grown with ALD contains hydrogen and can be used for passivation\cite{Dingemans2010}. We use this property of ALD-grown $\textrm{Al}_2\textrm{O}_3$ in a separate anneal step after $\textrm{Al}_2\textrm{O}_3$ deposition. We further postulate that the mechanical stability of our gate structure with pure Al is improved by encapsulation in this $\textrm{Al}_2\textrm{O}_3$ oxide. It is energetically more favorable to form $\textrm{Al}_2\textrm{O}_3$ than $\textrm{Si}\textrm{O}_2$ because the Gibbs free energy change of $\textrm{Al}_2\textrm{O}_3$ formation is lower than that of $\textrm{Si}\textrm{O}_2$. Al is known to reduce $\textrm{Si}\textrm{O}_2$ as confirmed by transmission electron microscopy of our samples\cite{Muller2015}.
It has also been shown that Al can spike through $\textrm{Si}\textrm{O}_2$ causing leakages and shorts to the channel\cite{Bierhals1998}.
Diffusivity of Al at temperatures around $300^\circ\text{C}$ can cause dewetting\cite{Thompson2012}, or void formation can occur due to different thermal expansion coefficients\cite{Hinode1989}.

We have found that the overlayer of 5~nm of $\textrm{Al}_2\textrm{O}_3$ alone in device type A stabilizes the pure aluminum electrodes. Without this overlayer void formation and dewetting-like behaviour was observed in all devices after the $300^\circ\text{C}$ annealing step (see Supplementary Information). The presence of an $\textrm{Al}_2\textrm{O}_3$ underlayer in device type B, between the $\textrm{Si}\textrm{O}_2$ and the Al gate structure could prevent the reduction reaction that would otherwise occur at the Al/$\textrm{Si}\textrm{O}_2$ interface. The under- and overlayer of $\textrm{Al}_2\textrm{O}_3$ respectively change the surface energy and suppress surface diffusivity of the Al film.

% We have measured transport characteristics and spectroscopy both before and after the annealing process at $T \simeq 4$ K and $T \simeq 10$ mK.

\begin{figure*}[!ht]
	\centering
	\includegraphics[width=\linewidth]{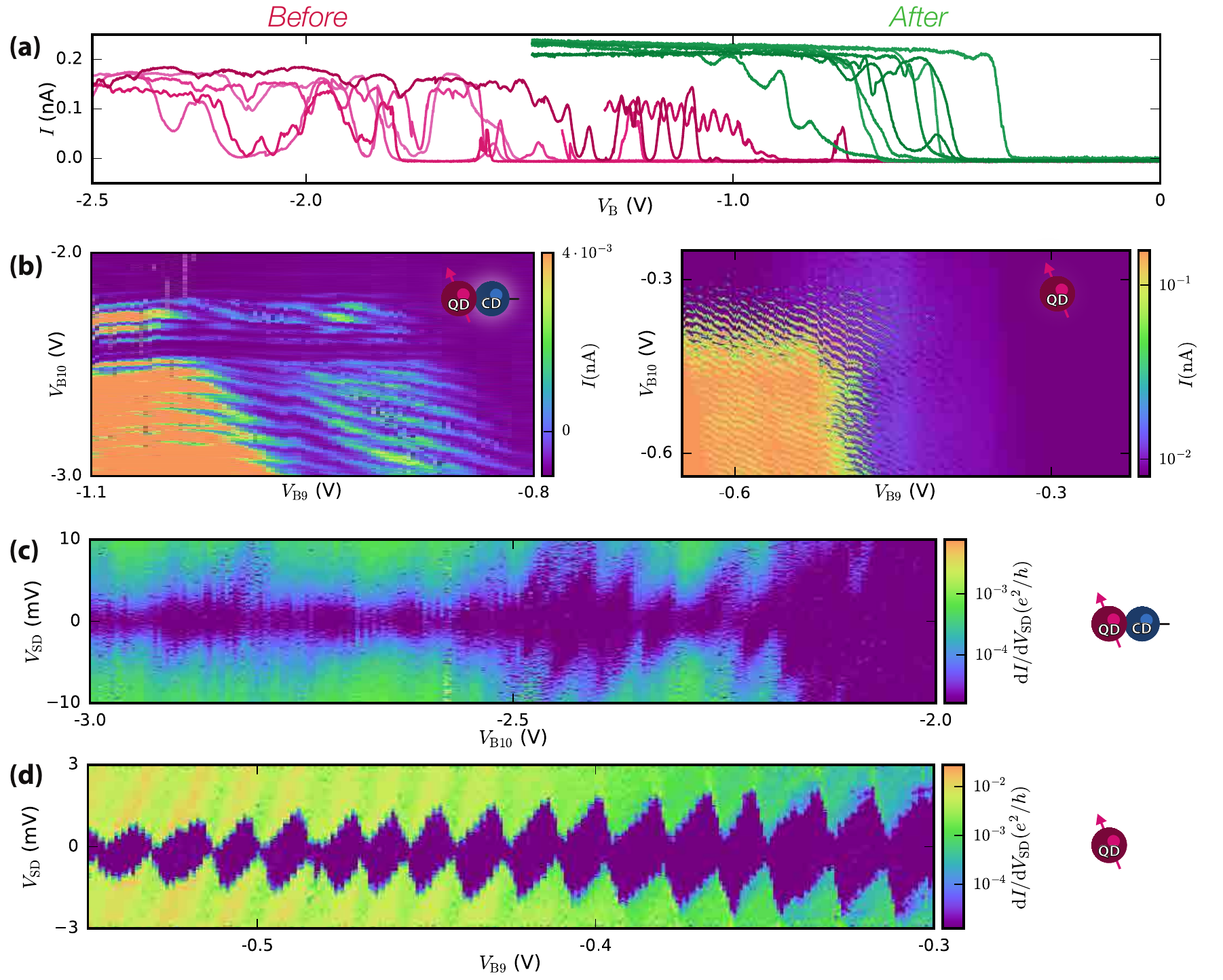}%
	\begin{minipage}{\linewidth}
		\subfloat{\label{fig:traces}
		}
	\end{minipage}%
	\begin{minipage}{.5\linewidth}
		\centering
		\subfloat{\label{fig:linear}
		}
	\end{minipage}%
	\begin{minipage}{\linewidth}
		\subfloat{
		\label{fig:bias_pre}
		}
	\end{minipage}%
	\begin{minipage}{\linewidth}
		\subfloat{
		\label{fig:bias_post}
		}%
	\end{minipage}
\caption{Transport measurements of device A before and after annealing. \protect\subref{fig:traces} Current-voltage relation at $T \simeq 4.2~\mathrm{K}$ before (red) and after (green) annealing. Individual barriers are swept to 0 V while all other gate voltages are maintained above the threshold voltage with $V_\mathrm{SD} = 1~\text{mV}$ and $V_\mathrm{L} = -4.6~\text{V}$ before, and $V_\mathrm{L} = -2.4~\text{V}$ after anneal.
\protect\subref{fig:linear} Linear transport measurements of current $I$ as a function of barrier gate voltages $V_\mathrm{B10}$ and $V_\mathrm{B9}$ before (left) and after \protect\subref{fig:linear} (right) annealing.
\protect\subref{fig:bias_pre} The differential conductance $dI/dV_{\mathrm{SD}}$, as a function of source-drain bias $V_{SD}$ and barrier gate voltage $V_{B10}$ before annealing, taken at $V_\mathrm{B9} = -950~\text{mV}$. Irregular Coulomb-diamond like features are visible with charging energies up to 10 meV. \protect\subref{fig:bias_post} The same measurement after annealing with $V_\mathrm{B10} = -350~\text{mV}$ and $T \simeq 10~\text{mK}$. Well-defined Coulomb diamonds with charging energies of 2-3 meV are visible, corresponding to an electrostatically defined Coulomb island 180 nm in length and 80 nm in width.}
\label{fig:three}
\end{figure*}

We have measured various devices and report here characteristic behaviour of two types of samples A and B.
First, we will focus on device A and the influence of charge defects on the formation of electrostatically defined QDs.
Device A (shown in Figures~\ref{fig:AFM} and \ref{fig:schematic_layout}) does not contain a layer of $\textrm{Al}_2\textrm{O}_3$ between the $\textrm{Si}\textrm{O}_2$ and the Al gate structures. For its transport characteristics we will focus on hole transport. Device A is measured once before $\textrm{Al}_2\textrm{O}_3$ overlayer growth and subsequent annealing step. After this passivation procedure the device is measured again. The current $I$ is measured at the drain while the drain is kept at ground. Additionally, all voltages are given with respect to ground.

Figure~\ref{fig:traces} shows the current-voltage relation of the individual barriers before and after annealing. For these measurements all electrodes except the barrier electrode in question are at sufficiently high voltage to allow conduction, while one of the barrier electrodes is swept to such a voltage that the barrier will become opaque and thus ``cut off'' the channel locally.
 Cut-off voltages decrease from a spread of -2~V to -1~V to a more narrow region of -1~V to -0.5~V. This is expected to be due to the reduction of a charge layer formed by charge defects present before annealing. The decreased number of resonances in the current $I$ indicates disorder in the device has decreased substantially after annealing.
% Also visible is a noted decrease in the voltage needed to go from $I_{SD} \simeq 0.2~\mathrm{nA}$ to $I_{SD} \simeq 0~\mathrm{nA}$.
The threshold voltage $V_\mathrm{Th}$, here defined as the voltage at which the entire channel becomes conducting when applying the same voltage to all gates, is also lowered after annealing from $V_\mathrm{Th} \simeq -4$~V to  $V_\mathrm{Th} \simeq -2.3$~V (not shown).

We have observed that nearly all resonances associated with the barriers disappear after $\textrm{Al}_2\textrm{O}_3$ growth and the subsequent annealing procedure. We have confirmed this in separate bias spectroscopy measurements where Coulomb diamonds of order $E_C \simeq 10~\text{meV}$ coupled exclusively to one barrier gate are no longer present after annealing.

Before annealing, Figure~\ref{fig:linear} (left) contains at least two sets of parallel lines of Coulombs peaks with distinct coupling to B9 and B10. One set of nearly horizontal lines runs from -2.4~V to -2.3~V parallel to the $V_{B9}$ axis. We ascribe these horizontal lines to a charge defect underneath B10\cite{Spruijtenburg1}. Another feature is seen in the increased tunnel current running vertically around $V_\mathrm{B9} = -900~\text{mV}$, indicating another feature coupled strongly to B9 with different periodicity from the one underneath B10.
A set of irregular diagonal lines is also visible. Thus, this feature is coupled to both B9 and B10, signifying a quantum dot centered between B9 and B10. However, these lines are interrupted and not continuous when intersecting the features associated with the defects, which implies a QD in series with at least 2 defects.
After annealing the same device shows very ordered linear transport data in Figure~\ref{fig:linear} (right) ($T \simeq 10~\mathrm{mK}$) at $V_\mathrm{SD} = 0.3~\text{mV}$ with uninterrupted diagonal lines equally coupled to B9, B10. We extract $\sfrac{C_{\mathrm{B10}}}{C_{\mathrm{B9}}} \simeq 1$, where $C_{\textrm{B}_i}$ is the capacitance of barrier gate i to the dot.
The absence of a charge sensor does not allow us to probe how many electrons are on the dot but the amount of visible charge transitions ($\simeq$ 50) and the lack of interruptions in the charge transition lines indicates the dot has a very controllable charge occupation over a large range of voltages.

% The regularity and single periodicity of an intentional quantum dot coupled equally to B9 and B10 is evident and is a strong indication that the created quantum dot is stable and not perturbed by any electrically active defects.

% The inset of figure~\ref{fig:linearpost} shows the same device after annealing, measured at $T \simeq 4.2~\text{K}$.  Because of thermal energy and the size of the dot (180 nm x 80 nm) the features are broadenened, however even at this temperature the regularity and single periodicity of an intentional quantum dot coupled equally to B9 and B10 is evident.

% This is a strong indication that the created quantum dot is stable and not perturbed by any electrically active defects.

We next perform bias spectroscopy in order to more accurately determine the ability to form quantum dots. A bias spectroscopy before annealing reveals several overlapping sets of Coulomb diamonds, see Figure~\ref{fig:bias_pre}. This is indicative of charge centers in series with the QD. Coulomb diamonds with a charging energy of $E_C \simeq 10~\text{meV}$ are visible as well as another set of diamonds with a charging energy 3-5 meV. After annealing, Figure~\ref{fig:bias_post} shows very regular Coulomb diamonds with a well-defined charging energy of $E_C \simeq 2~\text{meV}$.
Below $V_{\mathrm{B9}} = 400~\text{meV} $ the diamonds are closing. Around $V_{\mathrm{B9}} = -400~\text{mV}$ a band of Coulomb blockade, with $|V_\mathrm{SD}| < 0.3~\text{mV}$, causes the diamonds not to close. This additional blockade is not due to the dot itself but rather from blockade present further along the 3 micron long channel formed by the lead gate which we were unable to mitigate.
The many charge transitions and regular spacing of the diamonds indicates the low level of disorder in the device.

% %
% \begin{minipage}{.5\linewidth}
% 	\centering
% 	\subfloat{\label{fig:linearpost}
% 	% \includegraphics[width=83mm]{fig2_linear_post}
% 	}
% \end{minipage}

\begin{figure}[ht]
	% \raggedright
	\includegraphics[width=\linewidth]{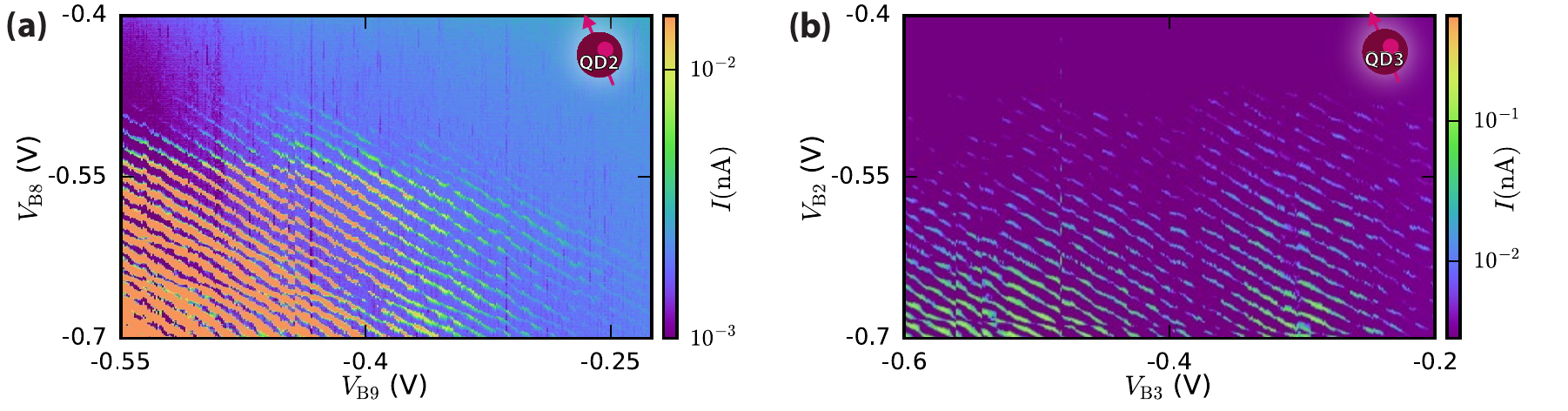}%
	\subfloat{
			\label{fig:fig3_left}
			}%
	\subfloat{
			\label{fig:fig3_right}
	}%
	\caption{Intentional quantum dots at both ends of device A.
	Linear transport measurements of current $I$ as a function of adjacent barrier gate voltages. An intentional quantum dot created \protect\subref{fig:fig3_left} between B2 and B3 with $V_\mathrm{L} = -2.9~\text{V}$, $V_\mathrm{SD} = 0.5~\text{mV}$ and \protect\subref{fig:fig3_right} between B8 and B9 with $V_\mathrm{L} = -3.1~\text{V}$, $V_\mathrm{SD} = 1~\text{mV}$. Diagonal current peaks indicate charge transitions coupled almost equally to both B2, B3 ($\sfrac{C_{\mathrm{B2}}}{C_{\mathrm{B3}}} \simeq 0.75$) and B8, B9 ($\sfrac{C_{\mathrm{B8}}}{C_{\mathrm{B9}}}  \simeq 0.95$).
	\label{fig:longquantumdots}
}

\end{figure}

Figure~\ref{fig:longquantumdots} shows additional linear transport measurements taken after annealing at two different ends of device A. Quantum dots QD2 and QD3, indicated in Figure~\ref{fig:one} are about 2 micron apart. Both Figure~\ref{fig:fig3_left} and Figure~\ref{fig:fig3_right} have the same regularly spaced charge transitions as in Figure~\ref{fig:linear} (right). The charge transitions have equal capacitive coupling to the barrier gates defining the quantum dot and there is no interference from unwanted charge defects.
%This indicates the annealing process allows us to reproducibly create quantum dots 180 nm in length and that the QD is no longer perturbed after passivation of the defects. 
We conclude from Figures~\ref{fig:three} and \ref{fig:longquantumdots} that our annealing process passivates the majority of defects, and allows formation of hole quantum dots of ~180 nm long.% $\textrm{P}_\textrm{b}$ centers.

\begin{figure}[ht]
	\includegraphics[width=\linewidth]{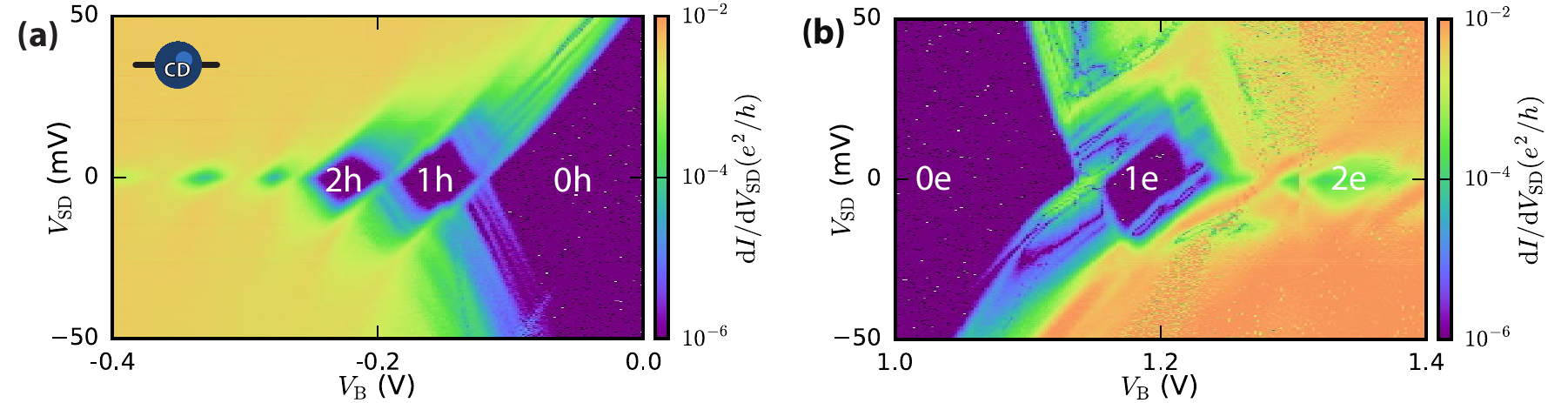}%
	\subfloat{
				\label{fig:amphoteric_hole}
			}%
	\subfloat{
			\label{fig:amphoteric_electron}
	}
	\caption{Amphoteric nature of the charge center in device B.\protect\subref{fig:amphoteric_hole} the hole operation regime ($V_\mathrm{L} = -1.6~\text{V}$), and the \protect\subref{fig:amphoteric_electron} electron operation regime ($V_\mathrm{L} = 1.7~\text{V}$) of the same device.
	Both the hole and electron operation regime show clear charge transitions, with regions of charge stability indicated with h and e for hole- and electron-occupancy, respectively. The charging energy of the first charge state of both electron and hole are found to be $E_C \simeq 10~\text{meV}$. }
	\label{fig:amphoteric}
\end{figure}

Now, we will investigate the nature of a single electrically active defect, left after the annealing procedure.
Device B, in contrast to device A, contains a layer of $\textrm{Al}_2\textrm{O}_3$ between the $\textrm{Si}\textrm{O}_2$ and the gate electrodes. We have observed that this further increases the yield of our devices. The specific effects of this interlayer on transport characteristics before and after annealing, on the same device, were not studied. 
After annealing we observe, in all devices with this interlayer, comparable threshold voltages and the same low probability of finding charge traps underneath individual barriers.

 In our ambipolar device structure, the same heterostructure and its associated imperfections can be probed in both the hole and electron operation regime. Figure~\ref{fig:amphoteric} shows a bias spectroscopy of a defect still present after annealing underneath a single barrier. The Coulomb diamonds near the valence band and the conduction band are very similar, and several charge states can be clearly observed. After the last charge transition ($1\textrm{h}\rightarrow 0\textrm{h}$ and $1\textrm{e}\rightarrow 0\textrm{e}$) the diamonds opens up and no charge transitions are observed up to a very high bias, for both the hole and electron operation regime. This indicates that the dot has been emptied completely, although without additional verification from e.g. a charge sensing experiment this cannot be said with absolute certainty. The charging energy on both the hole and electron side is found to be $E_C \simeq 10~\text{meV}$. 

%This is consistent with the known amphoteric nature of the $\textrm{P}_\textrm{b}$ center.

% In nearly all the devices we have measured we observe that if there is a charge transition near the valence band it coincides with a similar charge transition near the conduction band. This is consistent with the known amphoteric nature of the $\textrm{P}_\textrm{b}$ center.

We will now discuss 4 different explanations of these remaining charge transitions, namely dopant atoms, $\textrm{E}'$ centers, unintentional quantum dots, and $\textrm{P}_\textrm{b}$ centers.
%These features could be interpreted in several ways, which we will discuss now.

The chance that we are dealing with a dopant atom is small, since our background doping is $\leq 10^{12} \textrm{cm}^{-3}$. Additionally, these would not show amphoteric behaviour and charging energies below 20 meV.

$\textrm{E}'$ centers, another type of $\textrm{Si}\textrm{O}_2$ defect, are described as non-amphoteric deep hole traps and responsible for fixed positive charge.\cite{Kim1988} Primarily induced by radiation, $\textrm{E}'$ centers are often characterized as border traps\cite{Fleetwood}, located further from the conduction channel. Whether or not this defect could cause Coulomb blockade would depend on its distance from conduction channel. Radiation induced damage caused by electron beam evaporation\cite{Schiele}, used in our sample fabrication, could lead to higher incidence of this type of defect.  

One other explanation for the amphoteric nature of the observed charge transition is the formation of a so-called unintentional quantum dot, created by e.g. disorder or strain\cite{Thorbeck2014}. In case of the latter the conduction or valence band is modulated by strain, induced by the difference in thermal expansion coefficients between the silicon and the aluminum gates in the heterostructure. This can create tunnel barriers on each side of the barrier gate, resulting in an unintentional quantum dot. The fact that we observe quantum dot behaviour in a small percentage of barrier gates could be explained by variations in grain size of our constituent Al, resulting in non-uniform strain and tunnel barriers which vary from barrier to barrier.

A system very comparable to the $\textrm{P}_\textrm{b1}$ center is the single dangling bond on hydrogen-terminated intrinsic silicon, i.e. without an oxide. Density functional theory (DFT) calculations of this single dangling bond ($\textrm{D}^\textrm{0}$) have shown that the associated bound state has an energy level approximately 13~meV above the valence band maximum\cite{Schofield2013}. This agrees very well with our extracted value for the charging energy.

The multiple charge states visible in both the hole and electron operation regime preclude us from identifying the simple case of a single $\textrm{P}_\textrm{b}$ center with only three charge states: $\textrm{D}^-$, $\textrm{D}^0$, $\textrm{D}^+$.
The origin of the Coulomb diamonds and associated charge states can be explained by the non-singular nature of the defect: Two $\textrm{P}_\textrm{b}$ centers in parallel would also give rise to overlapping Coulomb diamonds, while the different addition energies of the neighbouring Coulomb diamonds would not be expected. It is also known that very proximate dangling bonds can form molecular type bonding, which could account for the observation of more than three charge states\cite{Schofield2013}.

The ability to resolve single $\textrm{P}_\textrm{b}$ centers could be attributed to the equilibrium density and the corresponding possibility of finding a small but definite number of unpassivated $\textrm{P}_\textrm{b}$ centers underneath a barrier. This corresponds to our findings that a finite number of barriers still shows an electrically active defect. To confirm this type of model, the device structure proposed in this article combined with statistical analysis on the incidence of charge defects with varying annealing processes would be needed. Electrical detection of spin resonance (EDSR) \cite{Xiao2004} would further provide strong experimental evidence to confirm if these remaining charge defects are caused by unpassivated $\textrm{P}_\textrm{b}$ centers.

% \ce{(P_bH)^0 + H+ -> P_b+ + H2}

% \begin{figure}[!ht]
% 	\subfloat[]{
% 			\includegraphics[width=83mm]{fig5_overview}
% 			\label{fig:singlepb_overview}
% 	}
%
%
% 	\subfloat[]{
% 				\includegraphics[width=30mm]{}
% 				\label{fig:singlepb_gfactor}
% 			}
% 	\subfloat[]{
% 				\includegraphics[width=30mm]{fig5_bdepend}
% 				\label{fig:singlepb_gfactor}
% 			}
% 	\subfloat[]{
% 			\includegraphics[width=30mm]{fig5_zoom}
% 			\label{fig:singlepb_zoom}
% 	}
% 	\caption[LoF]{gfactor of a single $\textrm{P}_\textrm{b}$ center. }
% 	\label{fig:singlepb}
%
% \end{figure}

In conclusion, we have demonstrated that an annealing process using ALD-grown $\textrm{Al}_2\textrm{O}_3$ is able to passivate the majority of the electrically active defects present at the Si/$\textrm{Si}\textrm{O}_2$ interface. We attribute this to the passivating properties of hydrogen present in the $\textrm{Al}_2\textrm{O}_3$ layer. After annealing the ability to form electrostatically defined quantum dots markedly improves with dot lengths of at least 180 nm. These quantum dots show many charge transitions, indicating the low level of disorder in the devices. We believe this is most likely due to the elimination of the majority of $\textrm{P}_\textrm{b}$ centers. 
The ambipolar device architecture then allows us to reveal the amphoteric behaviour of a remaining charge defect underneath a single barrier with energy levels $\simeq$10 meV above the valence band and below the conduction band. 

The ability to form singular quantum dots in this heterostructure is strongly dependent on the density of charge defects at the interface. The control of the spatial density of charge defects in Si planar quantum dots is thus necessary in order to succesfully and reproducibly scale up to many devices in a proposed quantum computer. For $\textrm{P}_\textrm{b}$ centers it is known that the passivation/depassivation reaction leads to an equilibrium, which implies there will always be a finite defect density at the Si/$\textrm{Si}\textrm{O}_2$ interface. 
In order to eliminate defects at this interface, another crystalline material to supplant the amorphous $\textrm{Si}\textrm{O}_2$ would be ideal. Candidates for this could either be crystalline dielectrics or other crystalline materials such as SiGe\cite{Borselli2014}.

The architecture presented here successfully avoids dewetting of Al in the annealing step by use of $\textrm{Al}_2\textrm{O}_3$, which simultaneously provides hydrogen for the passivation reaction. The ability of the ambipolar architecture to investigate the amphoteric nature of defects could be a very useful tool in the arsenal to further investigate charge defects in all manner of materials.

% Understanding the processes at the Si/SiO2 interface where the 2DEG/2DHG lives is one of the most important issues to be addressed in scaling of this architecture.

%\bibliography{/Users/waka/phd/publications/library}{}

\section*{Acknowledgements}
The authors would like to thank J.W. Mertens and A.A.I. Aarnink for the growth of the $\textrm{Al}_2\textrm{O}_3$, and J. Schmitz for valuable discussion and input. We acknowledge financial support through the EC FP7-ICT initiative under Project SiAM No 610637. W.G. van der Wiel and F.A. Zwanenburg acknowledge support from the Foundation for Fundamental Research on Matter (FOM), which is part of the Netherlands Organization for Scientific Research (NWO).

\section*{Author contributions statement}

P.C.S. conducted the experiment(s) and wrote the manuscript text, P.C.S., S.V.A., and F.M. fabricated the samples, P.C.S, S.V.A, and F.A.Z. analysed the results. All authors reviewed the manuscript.

\section*{Additional information}
All authors declare that there are no potential competing financial interests.

\end{document}